\documentclass[a4paper]{jpconf}
\usepackage{graphicx}
\usepackage{amsmath}
\usepackage{amsfonts}
\usepackage{amssymb}
\begin{document}
\title{Braided fermions from Hurwitz algebras}

\author{Niels G Gresnigt}

\address{Xi'an Jiaotong-Liverpool University, Department of Mathematical Sciences. 111 Ren'ai Road, Dushu Lake Science and Education Innovation District, Suzhou Industrial Park, Suzhou, 215123, P.R. China}

\ead{niels.gresnigt@xjtlu.edu.cn}

\begin{abstract}

Some curious structural similarities between a recent braid- and Hurwitz algebraic description of the unbroken internal symmetries for a single generations of Standard Model fermions were recently identified. The non-trivial braid groups that can be represented using the four normed division algebras are $B_2$ and $B_3^c$, exactly those required to represent a single generation of fermions in terms of simple three strand ribbon braids. These braided fermion states can be identified with the basis states of the minimal left ideals of the Clifford algebra $C\ell(6)$, generated from the nested left actions of the complex octonions $\mathbb{C}\otimes\mathbb{O}$ on itself. That is, the ribbon spectrum can be related to octonion algebras. Some speculative ideas relating to ongoing research that attempts to construct a unified theory based on braid groups and Hurwitz algebras are discussed.     

\end{abstract}

\section{Introduction}

Leptons and quarks are identified with representations of the gauge group $U(1)_Y\times SU(2)_L\times SU(3)_C$ in the Standard Model (SM) of particle physics. Despite its success in accurately describing and predicting experimental observations, this gauge group lacks a theoretical basis. Why has Nature chosen these gauge groups from an infinite set of Lie groups, and why do only some representations correspond to physical states? A second shortcoming of the SM is the lack of gravity, or equivalently, its unification with General Relativity (GR).

Preon models attempt do explain SM structure by proposing that leptons and quarks are not truly fundamental, instead being composed of a smaller number of subparticles called preons. The most famous of these is the 1979 Harari-Shupe preon model in which all SM leptons and quarks are composites of two spin half fermions \cite{Harari1979,Shupe1979}. This model is able to account for many of the observed SM symmetries but not without introducing some problems. For one, the preons must be confined but via what mechanism is not known. Attempts to account for preon confinement via a QCD-like confinement mechanism inevitably requires sacrificing the original model's simplicity.

In 2006, Bilson-Thompson proposed representing a single generation of SM fermions in terms of simple three stand ribbon braids \cite{Bilson-Thompson2005}. Inspired by the model of Harari and Shupe, it duplicates the success of that model while also avoiding some of its shortcomings \footnote{Other schemes have also been able to reproduce the structure of the Harari and Shupe preon model. One such example is a knot model by Finkelstein, in which the trefoil knots, understood as the $j=3/2$ irreducible representations of the quantum group $SU_q(2)$, form the basis of fermion states \cite{Finkelstein2007,Finkelstein2005}. In that scheme the Rishons of the Harari-Shupe model may be identified with the fundamental representations \cite{finkelstein2009knots,finkelstein2014preon}. A second example is the non-relativistic phase-space approach of $\dot{Z}$enczykowski \cite{Zenczykowski2007,Zenczykowski2007a}.}. The binding of preons is purely topological, and preons do not have their own independent existence. These braided objects find an embedding into Loop Quantum Gravity (LQG) when the cosmological constant is not zero \cite{Bilson-Thompson2007} \footnote{In such cases the underlying spin networks of LQG are $q$-deformed, giving a framing to its edges and requiring that these edges be labelled by representations of the quantum group $U_q(2)$ instead of its classical counterpart \cite{smolin2002quantum}.}. A braided description of matter therefore provides a potential means of unifying the SM with GR, with both being emergent from a theory of quantum gravity based on ($q$-deformed) spin networks. 

Dixon has demonstrated that many of the SM symmetries can be seen to arise from tensor products of Hurwitz algebras, in particular the algebra $T=\mathbb{C}\otimes \mathbb{H}\otimes\mathbb{O}$ \cite{dixon2013division,dixon1990derivation,dixon2010division}. Starting from a split algebra basis in a tensor product of Hurwitz algebras, say $\mathbb{C}\otimes\mathbb{O}$, one can define projection operators corresponding to primitive idempotents. These reduce $\mathbb{C}\otimes\mathbb{O}$ spinors into orthogonal subspaces, each of which transforms correctly as a generation of leptons and quarks under $SU(3)_C\otimes U(1)_Y$. In similar constructions by Furey and Stoica, the basis states of minimal left ideals of the Clifford algebra $C\ell(6)$ are shown to transform like a single generation of fermions under the unbroken symmetries $SU(3)_C\otimes U(1)_{EM}$ \cite{furey2016standard,stoica2017standard}. 

Unlike Lie and Clifford algebras, of which there are infinitely many, there are only four normed division algebras. Specific Clifford algebras are generated via the nested left actions of these algebras (or tensor products of them) on themselves. The bivectors of these close under commutation to give a representation of a specific Lie algebra. In this way, the Hurwitz algebras are very generative. It is well known that the complex numbers generate $U(1)$ and that the quaternions $\mathbb{H}$ are isomorphic to $SU(2)$. The automorphism group of the octonions is the exceptional Lie group $G_2$, which contains the subgroup $SU(3)$. The octonions have been studied in relation to the structure of quarks since 1973 \cite{gunaydin1973quark, gunaydin1974quark}. These observations suggest that the Hurwitz algebras may be the appropriate mathematical structures from which to derive the SM.

Several curious structural similarities between the braid and Hurwitz algebra descriptions of SM fermions have recently been identified \cite{gresnigt2017quantum2,gresnigt2018braids}. The complex numbers and quaternions contain representations of precisely the braid groups used to construct braided fermions in the Helon model. Furthermore, the twist and permutation structure of the ribbon braids in the Helon model coincides exactly with the states that span the minimal left ideals of $C\ell(6)$, generated from the nested left action of $\mathbb{C}\otimes\mathbb{O}$ on itself, indicating a relation between the braid spectrum and octonion algebras. These unexpected connections provide the first steps towards developing a unified theory based on braid groups and Hurwitz algebras. This paper reviews these recent results and comments of current research.

%%%%%%%%%%%%%%%%%%%%%%%%%%%%%%%%%%%%%%%%%%%%%%%%%%%%
\section{Hurwitz algebras}

\begin{figure}
\centering
\includegraphics[width=0.20\linewidth]{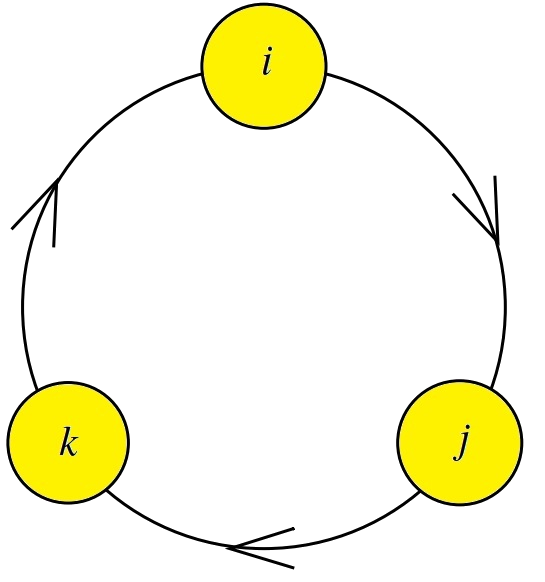}
\qquad\qquad\qquad
\includegraphics[width=0.3\linewidth]{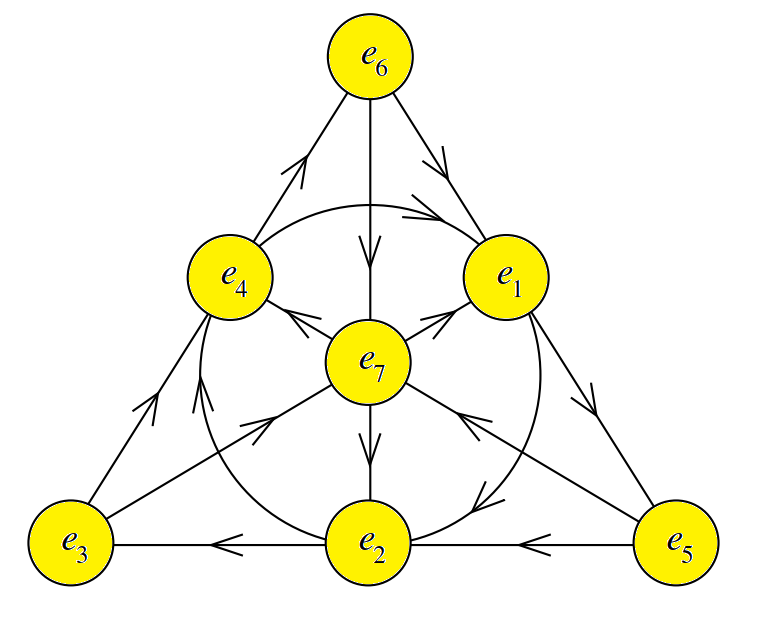}
\caption{Quaternion multiplication $I^2=J^2=K^2=IJK=-1$, and octonion multiplication represented using a Fano plane.}
\label{divisionalgebras}
\end{figure}

A Hurwitz algebra is a finite-dimensional unital (but not necessarily associative) algebra $A$ with a nondegenerate quadratic form $Q$ satisfying $Q(a)Q(b)=Q(ab)$. There exist two kinds of Hurwitz algebras; normed division algebras, of which there are exactly four ($\mathbb{R,C,H,O}$), and split algebras (split-$\mathbb{C,H,O}$) \footnote{A division algebra is an algebra over a field where division is always possible, with the exception of division by zero. In a split algebra on the other hand, there exists at least one $x\in A$ such that $Q^2(x)=xx^*=0$ where $Q(x)$ is the norm of $A$ defined in terms of a conjugation $x\mapsto x^*$. For a division algebra, the multiplicative inverse of an element $x$ is given by $x^*/Q^2(x)$.}. Generalizing from the real number $\mathbb{R}$ to the complex numbers $\mathbb{C}$, one gives up the ordered property of the reals. Generalizing in turn to the quaterions $\mathbb{H}$ one furthermore loses commutatativity. The quaternions are spanned by ${1,I,J,K}$ with $1$ being the identity and $I,J,K$ satisfying
\begin{eqnarray}
I^2=J^2=K^2=IJK=-1.
\end{eqnarray}
Finally, in moving to the octonions $\mathbb{O}$ one no longer has associativity. An excellent introduction to the octonions is given by Baez \cite{baez2002octonions}. The octonions are spanned by the identity $1=e_0$ and seven $e_i$ satisfying
\begin{eqnarray}
e_ie_j=-\delta_{ij}e_0+\epsilon_{ijk}e_k,\quad\textrm{where}\quad
e_ie_0=e_0e_i=e_i,\;e_0^2=e_0,
\end{eqnarray}
and $\epsilon_{ijk}$ is a completely antisymmetric tensor with value +1 when $ijk = 123,\;145,\;176,\;246,\;257,\\347,\;365$. The multiplication of quaternions and octonions is shown in Figure \ref{divisionalgebras}.

%%%%%%%%%%%%%%%%%%%%%%%%%%%%%%%%%%%%%%%%%%%%%%%%%%%%
\subsection{Clifford and Lie algebras from Hurwitz algebras}

Unlike Lie and Clifford algebras, there are only a finite number of Hurwitz algebras\footnote{We use the term Hurwitz algebra here instead of normed division algebra because, the reduction of spinor spaces considered by Dixon \cite{dixon2013division}, and the construction of minimal ideals considered by Furey \cite{furey2016standard} depend on the existence of nilpotents and primitive idempotents. These do not exist in normed division algebras, but do in their split algebra cousins. For example, when working in $\mathbb{C}\otimes\mathbb{H}$ or $\mathbb{C}\otimes\mathbb{O}$ one uses a primitive idempotent to select a split basis to work with.}. These algebras are very generative. Given a Hurwitz algebra identified with a spinor space, one acts on this spinor space with the algebra itself, from the left, right, or from both sides. Sometimes this does not make a difference, and sometimes it does. Considering all the repeated nested actions generate the adjoint algebras, which are specific Clifford algebras. The bivectors of a Clifford algebra close under commutation, providing a representation of the Lie algebra that preserves the metric of the associated pseudo-orthogonal space.

The small set of Hurwitz algebras thus provide a means of selecting (or rather a mechanism for generating) very specific Clifford and Lie algebras from an otherwise infinite sets. Particularly interesting is the case of the complex octonions. Although this algebra is not associative, the adjoint algebras $C\ell(7,0)\cong Cl(6)$ are. 
%%%%%%%%%%%%%%%%%%%%%%%%%%%%%%%%%%%%%%%%%%%%%%%%%%%
\section{Unbroken fermion symmetries from Hurwitz algebras}

%%%%%%%%%%%%%%%%%%%%%%%%%%%%%%%%%%%%%%%%%%%%%%%%%%%
\subsection{Primitive idempotents and orthogonal subspaces}

The extensive works of Dixon \cite{dixon2013division,dixon1990derivation,dixon2010division} focus on deriving the SM from tensor products of Hurwitz algebra. The approach is to reduce spinor spaces of Hurwitz algebras (as well as their associated action algebras) into sets of orthogonal subspaces. This is achieved by finding projection operators within the algebras, corresponding to primitive idempotents obtained from a resolution of the identity. In the case of $\mathbb{C}\otimes\mathbb{O}$ the projectors can be used to split the algebra into four orthogonal subspaces, each a $SU(3)$ multiplet. This is because the automorphism group that keeps $e_7$ fixed is $SU(3)$. Each spinor in $\mathbb{C}\otimes\mathbb{O}$ is resolved into $\mathbf{1}\oplus\mathbf{3}\oplus\mathbf{\overline{3}}\oplus\mathbf{\overline{1}}$ with respect to this chosen $SU(3)$ automorphism group\cite{dixon2013division}. The same algebra admits a $U(1)$ generator that produces the correct hypercharges. 
%%%%%%%%%%%%%%%%%%%%%%%%%%%%%%%%%%%%%%%%%%%%%%%%%%%
\subsection{Minimal left ideals of $C\ell(6)$}

The goal of identifying SM structure from tensor products of Hurwitz algebras is not unique to Dixon. More recently, Furey and Stoica have likewise looked at $\mathbb{C}\otimes\mathbb{O}$, and its associated left action algebra $C\ell(6)$, to identify unbroken SM symmetries \cite{furey2016standard,stoica2018leptons}. However, in their works the implementation of $C\ell(6)$ is significantly different. In particular, the spinors representing fermions in this scheme correspond to basis states of the minimal left ideals of $C\ell(6)$. 

Under a change of basis (induced by a projector), the generating space of the Clifford algebra $C\ell(2n)$ can be partitioned into two maximal totally isotropic subspaces (MTIS) from which two minimal ideals can be constructed. One can label the vectors of a split basis for $\mathbb{C}\otimes\mathbb{O}$ as
\begin{eqnarray}
\alpha_1 &=&\frac{1}{2}(-e_5+ie_4),\qquad \alpha_2=\frac{1}{2}(-e_3+ie_1),\qquad \alpha_3=\frac{1}{2}(-e_6+ie_2),\\
\alpha_1^{\dagger}&=&\frac{1}{2}(e_5+ie_4),\qquad \alpha_2^{\dagger}=\frac{1}{2}(e_3+ie_1),\qquad \alpha_3^{\dagger}=\frac{1}{2}(e_6+ie_2),
\end{eqnarray}
which also serves as a basis for $C\ell(6)$. It is readily checked that $\alpha_i$ and $\alpha_i^{\dagger}$ are nilpotents, and furthermore behave as a set of fermionic creation and annihilation operators, satisfying
\begin{eqnarray}
\lbrace \alpha_i,\alpha_j \rbrace=0,\; \lbrace \alpha_i^{\dagger},\alpha_j^{\dagger} \rbrace=0,\; \lbrace \alpha_i^{\dagger},\alpha_j \rbrace=\delta_{ij}.
\end{eqnarray}
The $\lbrace\alpha^{\dagger}_i\rbrace$ span one three complex-dimensional MTIS. It is readily checked that $\alpha_i\omega=0$. A minimal left ideal is then given by allowing $C\ell(6)$ to act on the primitive idempotent $\omega\omega^{\dagger}=\alpha_1\alpha_2\alpha_3\alpha_3^{\dagger}\alpha_2^{\dagger}\alpha_1^{\dagger}$. The first ideal is then
\begin{eqnarray}\label{ideal}
\nonumber S^u\equiv &{}&\\
\nonumber &{}&\;\;\nu \omega\omega^{\dagger}+\\
\nonumber \bar{d}^r\alpha_1^{\dagger}\omega\omega^{\dagger} &+& \bar{d}^g\alpha_2^{\dagger}\omega\omega^{\dagger} + \bar{d}^b\alpha_3^{\dagger}\omega\omega^{\dagger}\\
\nonumber u^r\alpha_3^{\dagger}\alpha_2^{\dagger}\omega\omega^{\dagger} &+& u^g\alpha_1^{\dagger}\alpha_3^{\dagger}\omega\omega^{\dagger} + u^b\alpha_2^{\dagger}\alpha_1^{\dagger}\omega\omega^{\dagger}\\
&+& e^{+}\alpha_3^{\dagger}\alpha_2^{\dagger}\alpha_1^{\dagger}\omega\omega^{\dagger},
\end{eqnarray}
where $\nu$, $\bar{d}^r$ etc. are suggestively labeled complex coefficients. The complex conjugate system analogously gives a second linearly independent minimal left ideal.

%%%%%%%%%%%%%%%%%%%%%%%%%%%%%%%%%%%%%%%%%%%%%%%%%%%
\section{Representation of fermions as three strand ribbon braids}
%%%%%%%%%%%%%%%%%%%%%%%%%%%%%%%%%%%%%%%%%%%%%%%%%%%%
\subsection{The Helon model}

\begin{figure}[h!]
\centering
 \includegraphics[scale=0.25]{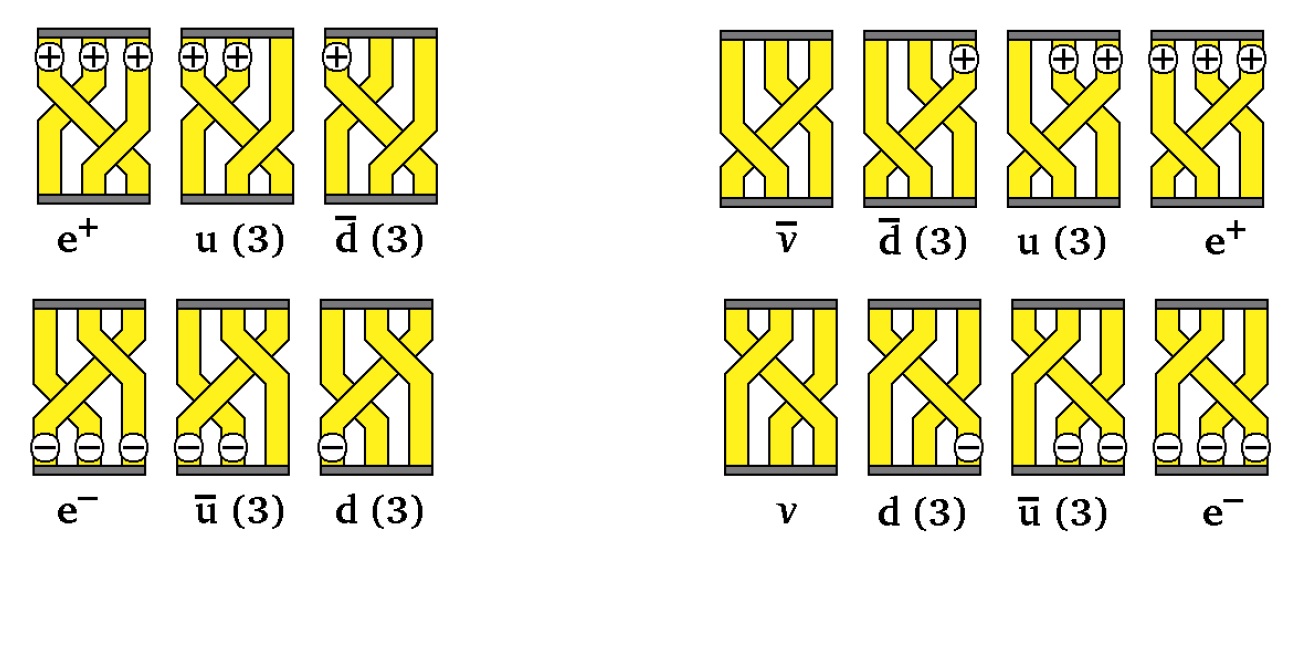}
\caption{The Helon model of Bilson-Thompson in which the first generation SM fermions are represented as braids of three (possibly twisted) ribbons. The right hand side of the figure shows the left-handed particles and right-handed antiparticles partaking in the weak interaction.  Used with permission. Source, \cite{Bilson-Thompson2005}.}
\label{helonmodel}
\end{figure}

Bilson-Thompson's Helon model maps the simplest ribbon braids, consisting of three twisted ribbons (called helons) and two crossings to the first generation of SM fermions \cite{Bilson-Thompson2005}. Quantized electric charges of $\pm e/3$ are represented by integral twists of $\pm 2\pi$ on the ribbons of the braids. These ribbons are combined into triplets by connecting the tops and bottoms of three ribbons to a parallel disk. The color charges of quarks are accounted for by the permutations of twists on certain braids, and simple topological processes are identified with the electroweak interaction, the color interaction, and conservation laws. The representation of first generation SM fermions in terms of braids is shown in Figure \ref{helonmodel}.

%%%%%%%%%%%%%%%%%%%%%%%%%%%%%%%%%%%%%%%%%%%%%%%%%%%
\subsection{Braid group representations from Clifford algebras}

The Artin braid group on $n$ strands is denoted by $B_n$ and is generated by elementary braids $\left\lbrace \sigma_1,...,\sigma_{n-1}\right\rbrace$ subject to the relations
\begin{eqnarray}\label{braidrelations}
\sigma_i \sigma_j=\sigma_j\sigma_i,\;\textrm{whenever}\; \vert i-j \vert > 1,\quad \sigma_i\sigma_{i+1}\sigma_i=\sigma_{i+1}\sigma_i \sigma_{i+1},\;\textrm{for}\; i=1,....,n-2.
\end{eqnarray}
The braid groups $B_n$ are an extension of the symmetric groups $S_n$ with the condition that the square of each generator being equal to one lifted. It can be shown that Clifford algebras admit representations of certain braid groups \cite{kauffman2016braiding}.

For a Clifford algebra $C\ell(n,0)$ over the real numbers generated by linearly independent elements $\left\lbrace e_1,e_2,...,e_n\right\rbrace $ with $e_k^2= 1$ for all $k$ and $e_ke_l=-e_le_k$ for $k\neq l$, the algebra elements $\sigma_k=\frac{1}{\sqrt{2}}(1+e_{k+1}e_k)$ form a representation of the circular\footnote{A circular braid on $n$ strings has $n$ strings attached to the outer edges of two circles which lie in parallel planes in $R^3$.} Artin braid group $B^c_n$. This means that the set of braid generators $\left\lbrace  \sigma_1,\sigma_2,...,\sigma_n\right\rbrace $, where
\begin{eqnarray}
\sigma_k = \frac{1}{\sqrt{2}}(1+e_{k+1}e_k),\;\textrm{whenever}\; 1\leq k <n,\qquad
\sigma_n =\frac{1}{\sqrt{2}}(1+e_1e_n),
\end{eqnarray}
satisfy the braid relations (\ref{braidrelations}). Although \cite{kauffman2016braiding} assumes that $e_k^2=1$ for all $k$, the proof likewise holds when $e_k^2=-1$, as is easily checked. 

%%%%%%%%%%%%%%%%%%%%%%%%%%%%%%%%%%%%%%%%%%%%%%%%%%%
\subsection{Braid group representations from Hurwitz algebras}

Using known isomorphisms between division algebras and Clifford algebras, one can investigate the braid content of (hyper)complex division algebras. The complex number algebra $\mathbb{C}$ is isomorphic to the real Clifford algebras $C\ell(0,1)\cong C\ell^+(2,0)$, where the $+$ indicates the even subalgebra of a Clifford algebra. A representation of $B_2$ in terms of $\mathbb{C}$ is given by
\begin{eqnarray}\label{complexgenerators}
\sigma_1=\frac{1}{\sqrt{2}}(1+i),\qquad \sigma_1^{-1}=\frac{1}{\sqrt{2}}(1-i),
\end{eqnarray} 
This representation is not faithful, with the order of the single generator (and its inverse) being eight. The inverse generator is simply the complex conjugate.

Moving on to the quaternions $\mathbb{H}$, one can use the isomorphism $\mathbb{H}\cong C\ell(0,2)\cong C\ell^+(3,0)$ to find a quaternionic representation of the braid group $B^c_3$
\begin{eqnarray}\label{quaterniongenerators}
\sigma_1&=\frac{1}{\sqrt{2}}(1+I),\;\;\sigma_2=\frac{1}{\sqrt{2}}(1+J),\;\;\sigma_3=\frac{1}{\sqrt{2}}(1+K).
\end{eqnarray}
The inverse braid generators are obtained via the quaternions conjugate.

The octonions are not associative and therefore they are not isomorphic to any Clifford algebra. The algebra is however alternative, and so any two basis elements $e_a$ do generate an associative algebra isomorphic to a quaternion subalgebra of $\mathbb{O}$. Thus, within $\mathbb{O}$ one finds seven distinct copies of $B_3^c$ (as well as seven distinct copies of $B_2$ generated from $e_a$).

%%%%%%%%%%%%%%%%%%%%%%%%%%%%%%%%%%%%%%%%%%%%%%%%%%%
\section{Relating Hurwitz algebras to the Helon model}

Considering the edges of each ribbon in the Helon model, one can identify this twisting with a braid in $B_2$. This braid group is represented, albeit not faithfully, by $\mathbb{C}$. Likewise, the braiding of three ribbons gives a braid in $B_3$. Because the ribbons are connected together at the top and bottom to a parallel disk, the relevant braid group is actually the circular version $B_3^c$. This is precisely the braid group that can be represented (again, not faithfully) using $\mathbb{H}$, or a quaternionic subalgebra of $\mathbb{O}$. The braid groups that can be represented using the Hurwitz algebras turn out to be precisely those needed to construct the Helon model. This is not the only hint at a deep connection between the two models. Identifying the ladder operators $\alpha_i$ and $\alpha_i^{\dagger}$ in \cite{furey2016standard} with braids in $B_3^c$, the basis states of the minimal left ideals of the complex octonions may be identified with the twist structure in the Helon model.
%%%%%%%%%%%%%%%%%%%%%%%%%%%%%%%%%%%%%%%%%%%%%%%%%%%%%%%%%%%%%%%%%%%%%%%%%%%%%%%%%%%%%
\subsection{Helon braids as the basis states of minimal left ideals of $C\ell(6)$}\label{braidedbasiststates}

In \cite{Bilson-Thompson2009} it was shown that braiding of ribbons and twisting of ribbons within three strand ribbon braids are interchangeable, and that any orientable\footnote{Orientable here meaning that the surface bounded by the ribbon braid is orientable.} ribbon braid can always be written as a vector $[a,b,c]$ where $a,b,c$ are all integers or all half integers, specifying the number of twists on each ribbon. This vector is a topological invariant. In particular, the generators of $B^c_3$ can be written as twist vectors as follows:
\begin{eqnarray}\label{braidgenerators}
\sigma_1\rightarrow \left[ 1/2,1/2,-1/2\right] ,\qquad \sigma_2\rightarrow \left[ -1/2,1/2,1/2\right],\qquad \sigma_3\rightarrow \left[ 1/2,-1/2,1/2\right],
\end{eqnarray}
with the inverse braid generators giving the negatives of these twist vectors. Care must be taken to account for the permutations induced by braidings when systematically exchanging braiding for twisting \cite{Bilson-Thompson2009}. It can likewise be shown that any orientable ribbon braid may be manipulated such that all the twisting of the ribbons is removed, exchanged for braiding \cite{gresnigt2018knotted}. One may write this as $[0,0,0]B=B[0,0,0]$ where $B\in B_3^c$.

Because the ribbon braids in the Helon model are orientable, they can be written purely as braid words in $B_3^c$. Letting $B=(\sigma_2^{-1}\sigma_1)$, we have for example
\begin{eqnarray}
\nonumber\nu &\rightarrow&  [0,0,0]B,\;\;e^+\rightarrow  [1,1,1]B=[0,0,0](\sigma_2\sigma_3)(\sigma_3\sigma_1)(\sigma_1\sigma_2)B,\\
\bar{d}^r &\rightarrow & [0,0,1]B=[0,0,0](\sigma_1\sigma_2)B,\;\;u^r \rightarrow  [0,1,1]B=[0,0,0](\sigma_2\sigma_3)(\sigma_3\sigma_1)B.
\label{positivestates}
\end{eqnarray}
If one now makes the identifications
\begin{eqnarray}\label{alphadagger}
(\sigma_1\sigma_2)\mapsto\alpha_1^{\dagger},\;\;(\sigma_3\sigma_1)\mapsto\alpha_2^{\dagger},\;\;(\sigma_2\sigma_3)\mapsto\alpha_3^{\dagger},\;\;\textrm{together\;with}\;\; \sigma_2^{-1}\sigma_1\mapsto\omega\omega^{\dagger},
\end{eqnarray}
and substitutes into the braid only representations of fermions, such as in equation (\ref{positivestates}), then the minimal left ideal $S^u$ of $C\ell(6)$, as in eqn., (\ref{ideal}) is recovered\footnote{The action of the basis states in the ideal is on the identity $[0,0,0]$ from left to right. Thus for example,
\begin{eqnarray}
\nonumber u^g &\rightarrow & [1,1,0](\sigma_2^{-1}\sigma_1)=[0,0,0](\sigma_1\sigma_2)(\sigma_2\sigma_3)(\sigma_2^{-1}\sigma_1)=[0,0,0]\alpha_1\alpha_3\omega^{\dagger}\omega.
\end{eqnarray}}. One may similarly reconstruct the second ideal for antiparticles in a similar manner. More details of this construction are given in \cite{gresnigt2018braids}.

Notice that if one is only concerned with the unbroken symmetries, and hence ignores the underlying braiding in Helon model braids (that is, ignore $\sigma_2^{-1}\sigma_1$ and instead take $\omega\omega^{\dagger}=I$ where $I$ is the identity braid), then one of the ideals consist entirely of positive braids (that is braid words made up of only braid generators), and the other ideal consists of negative braids made up entirely of inverse braid generators. 

%%%%%%%%%%%%%%%%%%%%%%%%%%%%%%%%%%%%%%%%%%%%%%%%%%%%%%%%%%%%%%%%%%%%%%%
\section{Extending the connections between the braid and Hurwitz algebraic models}
We now turn to some more speculative ideas to see if the connections between the braid- and Hurwitz algebraic description of SM symmetries can be further extended. In particular we are interested in investigating what the lack of faithfulness of the $\mathbb{C}$ and $\mathbb{H}$ representations of braid groups corresponds to physically, as well as trying to extend both the braid- and Hurwitz algebraic models to exactly three generations.

%%%%%%%%%%%%%%%%%%%%%%%%%%%%%%%%%%%%%%%%%%%%%%%%%%%%%%%%%%%%%%%%%%%%%%
\subsection{Electric charge structure of fermions from the $\mathbb{C}$ representation of $B_2$}

One speculative idea is that the lack of faithfulness of the representations of the braid groups $B_2$ and $B_3^c$ using $\mathbb{C}$ and $\mathbb{H}$ respectively naturally places a restriction on the number of distinct braided physical states that can be represented. The Helon model has no mechanism in place to restrict the number of allowable physical states. One could add more twisting to the ribbons and increase the complexity of braiding without limit, to get an infinite number of distinct states. Indeed, the model has been extended to a infinite number of generations \cite{Bilson-Thompson2008}. However, because the $\mathbb{C}$ and $\mathbb{H}$ representations of $B_2$ and $B_3^c$ representations are not faithful, many distinct braids are represented by the same (hyper)complex number. Each (hyper)complex number therefore represents an equivalence class of braids, each to be identified with a physical state. The finite set of equivalence classes then places a limit on the number of admissible physical states.

The twisted ribbons (helons) of the Helon model may be represented as braids in $B_2$ as 
\begin{eqnarray}
H_+=\sigma_1^2,\qquad H_-=\sigma_1^{-2},\qquad H_0=I,
\end{eqnarray} 
where $I$ denotes the identity unbraid. Because the braid word for each twisted ribbon is even, we therefore restrict ourselves to braids composed of an even number of braid generators\footnote{One may ask, why not identify a $\pm\pi$ twist with a charge of $\pm e/3$ instead? In that case the braid words would be odd instead. The problem is that when combining three such ribbons together, all the quarks would correspond to unorientable braids.}.

The $\mathbb{C}$ representation of $B_2$ is not faithful, dividing this braid group into eight equivalence classes, each represented by a complex number. It is readily checked that any even product of $B_2$ braid generators defined in (\ref{complexgenerators}) is equal to one of four complex numbers $\pm 1, \pm i$, each defining an equivalence class of braids.

The shortest possible braid in each class can be used as a representative for the equivalence class
\begin{eqnarray}
[I]\rightarrow 1,\qquad [\sigma_1^2]\rightarrow i,\qquad [\sigma_1^4]=[\sigma_1^{-4}]\rightarrow -1,\qquad [\sigma_1^{-2}](=[\sigma_1^{6}])\rightarrow -i.
\end{eqnarray}
The first, second, and fourth equivalence class representatives correspond to the helons $H_0$, $H_+$, and $H_-$ respectively, or equivalently, electric charges of $0$, $e/3$, and $-e/3$. Because $\sigma_1^4$ and $\sigma_1^{-4}$ both belong to the same equivalence class, it is impossible to represent both a ribbon with charge $2e/3$ and a ribbon with charge $-2e/3$ in a consistent manner\footnote{It is assumed here that every particle has a corresponding antiparticle. If one constructs a particle using an electric charge of $2e/3$ on one or more ribbons, then the corresponding antiparticle cannot be constructed because $\sigma_1^{-4}$ lies in the same equivalence class as $\sigma_1^4$.}. In other words, electric charges (for a single ribbon) outside the range $\left\lbrace 0, e/3,-e/3\right\rbrace $ are naturally excluded as a result of the complex representation of $B_2$ not being faithful.

%%%%%%%%%%%%%%%%%%%%%%%%%%%%%%%%%%%%%%%%%%%%%%%%%%
\subsection{Equivalence classes of braids from $\mathbb{H}$}

The braid generators in (\ref{quaterniongenerators}) and their products correspond to unit quaternions, which we write as $a+bI+cJ+dK$ with $a^2+b^2+c^2+d^2=1$. It is not difficult to show that any even product of braid generators is of the form 
\begin{eqnarray}\label{Hequivalence}
b_n\mapsto a+bI+cJ+dK,\quad \textrm{where} \quad a,b,c,d\in \left\lbrace 0, \pm \frac{1}{2}, \pm 1 \right\rbrace. 
\end{eqnarray}
This gives $2\times 4+2^4=24$ possibilities, but includes braids that lead to charge mixing, such as $\sigma_1^2,\sigma_2^2$, and $\sigma_3^2$. Written in pure twist form, $\sigma_1^2=[1,1,-1]$, $\sigma_2^2=[1,-1,1]$, and $\sigma_3^2=[-1,1,1]$. Since
\begin{eqnarray}
\sigma_1^2 \mapsto I,\qquad \sigma_2^2 \mapsto J,\qquad \sigma_3^2\mapsto K,
\end{eqnarray}
these quaternions must be ruled out from representing physical states. This leaves 18 equivalence classes, 16 with four terms and 2 with one term. It is not yet clear if it if possible to, and if so how, to identify 16 of these 18 equivalence classes with 16 distinct fermions, nor how to interpret the remaining two equivalence classes in this case.

%%%%%%%%%%%%%%%%%%%%%%%%%%%%%%%%%%%%%%%%%%%%%%%%%%%
\subsection{From one generation to exactly three}

There are several possibilities for extending the braid- and Hurwitz algebraic description from a single generation to exactly three generations. The algebra $C\ell(6)$ is large enough to represent three generations of fermions that transform as required under the unbroken symmetries \cite{furey2014generations}. It is however not yet clear how this three generation model relates to the single generation model based on minimal ideals. 

A possible answer may be provided by consider the exceptional Jordan algebra. Jordan algebras arose from Jordan's 1932 attempt to reformulate quantum mechanics by identifying the minimal axioms that should be satisfied by an algebra of observables \cite{mccrimmon2006taste}. The classification of Jordan algebras was worked out soon after by Jordan, Wigner, and Von Neumann, and showed that all but one Jordan algebra can be represented by hermitian matrices over $\mathbb{C}$ or $\mathbb{H}$. The one curious exception is the exceptional Jordan algebra $J_3(\mathbb{O})$. An element of $J_3(\mathbb{O})$ consists of three octonions and three real numbers, and can be written as a $3\times 3$ hermitian octonionic matrix
\begin{eqnarray}
\begin{pmatrix}
\alpha_1 & O_3 & \overline{O}_2 \\
\overline{O}_3 & \alpha_2 & O_1 \\
O_2 & \overline{O}_1 & \alpha_3
\end{pmatrix},\qquad \alpha_i\in\mathbb{R},\quad O_i\in \mathbb{O},
\end{eqnarray}
and where the bar indicates octonionic conjugation. This 27 dimensional exceptional Jordan algebra is the only Jordan algebra without a realization in terms of associative matrices. 

One might speculate that if the unbroken SM symmetries of one generation of fermions can be described by a single copy of $\mathbb{O}$ (or rather a copy of split $\mathbb{O}$ inside $\mathbb{C}\otimes\mathbb{O}$), then three generations of fermions requires three copies. It then seems natural to embed these three copies inside the exceptional Jordan algebra. This provides an explanation why there are no more than three generations, because for $n\geq 4$, $J_n(\mathbb{O})$ are not Jordan algebras. This idea has been explored recently by \cite{todorov2018deducing}. There it is shown that the gauge group of the SM can be obtained via the Borel-de Siebenthal theory, which describes the maximal closed connected subgroups of a compact Lie group that have maximal rank. 

The automorphism group of the exceptional Jordan algebra is the 52-dimensional exceptional Lie group $F_4$. Although the rank 4 group $F_4$ does not contain $SU(3)\times SU(2)\times U(1)$ as a subgroup, the subgroup of automorphisms which leave the three diagonal real elements in addition to one of the imaginary elements of each off-diagonal octonion invariant is exactly $SU(3)$ \cite{farnsworth2015standard}. The reduction from $F_4$ to $SU(3)$ therefore provides three $SU(3)$ singlets, three triplets, three antitriplets, and three antisinglets. One can likely extend this symmetry to $U(3)$ by finding a $U(1)$ generator as is done for an individual copy of the octonions \cite{furey2018three,dixon2013division}.

It remains to be shown how the ideas in \cite{todorov2018deducing} based on $J_3(\mathbb{O})$ relate to a construction of three families of fermions based on minimal left ideals. This would offer some insight into what an extension of the Helon model to exactly three generations might look like.

%%%%%%%%%%%%%%%%%%%%%%%%%%%%%%%%%%%%%%%%%%%%%%%%%%%%%
\section{Discussion}

%Unlike Lie algebras, Jordan algebras, and Clifford algebras, there are only a finite number of Hurwitz algebras. The four Hurwitz algebras are intimately connected to the Jordan and Clifford algebras, something which is apparent from a study of the representation theory of the latter algebras. The left and right actions of Hurwitz algebras on themselves (and more importantly tensor products of them) generate certain Clifford algebras. The bivectors of these close under commutation to give a representation of a Lie algebra. Work going back to the 1970s has established a connection between the Hurwitz algebras, in particular the octonions, and the internal symmetries of SM fermions. In a recent flurry of activity, these earlier works have been extended. 

There exist several curious structural similarities between recent braid- and Hurwitz algebraic descriptions of a single generation of fermions. Clifford algebras that are isomorphic to complex numbers and quaternions admit precisely those braid group representations from which the Helon model is contructed \cite{gresnigt2017quantum2}. Furthermore, the ribbon braids representing fermions in that model coincide exactly with the states that span the minimal left ideals of the adjoint algebra of the complex octonions, shown by Furey to describe one generation of leptons and quarks with unbroken $SU(3)_{C}$ and $U(1)_{EM}$ symmetry. 

A braided description of fermions offers a potential means of unifying fundamental matter with quantum spacetime in LQG. Taken collectively therefore, it might be hoped that Hurwitz algebras together with a braided interpretation of fermions provides an avenue for unifying spacetime with the SM, as well as providing a theoretical basis for the observed SM symmetries. 

Current ongoing research is looking at what the lack of faithfulness of the complex and quaternionic representations of $B_2$ and $B_3^c$ means physically. There is some indication that this lack of faithfulness restricts the possible electric charges of observed particles, and restricts the number of distinct physical states for a single generation. Another challenge is to extend both the braid- and Hurwitz algebra models to exactly three generations. One promising approach is to embed three copies of the octonions into the exceptional Jordan algebra, although many details remain to be worked out. 

%%%%%%%%%%%%%%%%%%%%%%%%%%%%%%%%%%%%%%%%%%%%%%%%%%%%%
\ack
This work is supported by the National Natural Science Foundation of China grant 11505143.

\section*{References}
%%%%%%%%%%%%%%%%%%%%%%%%%%%%%%%%%%%%%%%%%%%%%%%%%%%%%%%%%%%%%%%%%%%%%
\bibliography{NielsReferences}  % Replace xxx by your  usercode (no extension)
\bibliographystyle{unsrt}  
%%%%%%%%%%%%%%%%%%%%%%%%%%%%%%%%%%%%%%%%%%%%%%%%%%%%%%%%%%%%%%%%%%%%%

\end{document}